\documentclass[aps,subeqns.floatfix,twocolumn,amsmath,showpacs]{revtex4-1}
\usepackage{graphicx}
\usepackage{subfigure}
\usepackage{hyperref}

\begin{document}
\title{Bursting dynamics in a population of oscillatory and excitable Josephson junctions}
\author{ Chittaranjan Hens$^{1,2}$, Pinaki Pal$^3$, Syamal K. Dana$^1$}
\affiliation{$^1$CSIR-Indian Institute of Chemical Biology, Kolkata 700032, India}
\affiliation{$^2$Department of Mathematics, Bar-Ilan University, Ramat Gan, Israel}
\affiliation{$^3$Department of Mathematics, National Institute of Technology, Durgapur 713209, India}
\date{\today}

\begin{abstract}
 We report parabolic bursting in a globally coupled network of mixed population of oscillatory and excitable Josephson junctions. The resistive-capacitive shunted junction (RCSJ) model of  the superconducitng device is used for this study.  We focus on the parameter regime of the junction where its dynamics is governed by the saddle-node  on invariant circle (SNIC) bifurcation. In this SNIC regime, the bursting appears in a broad paramater space of the ensemble of mixed junctions. For a coupling value above a threshold, the network splits into two synchronized clusters when a reductionism approach is applied to reproduce the bursting behavior of the large network. The excitable junctions effectively induces a slow dynamics {\bf in the network to generate  bursting}. This  bursting is a generic property of a globally coupled network with a mixed population of dynamical nodes where each node posseses the SNIC property.

\end{abstract}

 \maketitle

\section{Introduction}
The superconducting Josephson junction shows self-oscillation \cite{levi, sastry, strogatz} for an applied constant current above a critical value. It is usually modeled as a resistive-capacitive-shunted junction (RCSJ) which  has its mechanical analog in a damped pendulum with a constant torque. A RCL-shunted junction (resistive-capacitive-inductively-shunted junction)  model  \cite{lobb, strogatz1, dana, kurt, pikovsky} was also used to include an inductive loading effect in an array of junctions where more complex dynamics   including chaos was seen. Interestingly, the superconducting device shows some typical spiking and bursting behaviors \cite{dana1, lynch} most commonly seen in a Type I  excitability  neuron \cite{izhikevich}. The bursting dynamics was also found prominent \cite{dana, dana1, crotty} in a periodically forced junction. This is due to the intrinsic SNIC characteristic of the junction in a selected parameter space \cite{strogatz, dana1, mackay}, which typically governs a class of bursting dynamics in Type I excitability neurons. 
\par Spiking is a repeatative firing state and bursting is a state of recurrent switching between a firing state or oscillatory state and a resting state. %
The minimal condition for  bursting in a system  necessitates the presence of an intrinsic slow-fast dynamics \cite{izhikevich, hindmarsh, ermentrout, rinzel}. As  example, in biological neurons, the simplest ionic processes  involved in spiking are due to the flow of $Na^{+}$ and $K^{+}$ ions across the
cell membrane, while the bursting may be observed when the fast spiking (FS) is controlled by a slow process like
$Ca^{++}$-gated $K^{+}$ ion movement across the membrane. The slow dynamics controls the firing or start of the oscillation and intermittently stops it when the trajectory of the dynamics moves slowly towards a steady state.  Alternatively, an excitable system when coupled to an oscillatory system, was found \cite{satya} to induce  a slow dynamics and thereby originates a type of chaotic bursting. 
 \par On a different context, a  mixed population of globally coupled inactive or excitable and active or oscillatory units was  investigated earlier \cite{daido,pazo, osipov, sitabhra, so-lucke}  in search of synchrony and  global oscillation. Such a global oscillation is practically important, particularly, in the context of a desired synchrony of the pacemaker cells \cite{jalife,kalsbeek}.  It is also important to know, in the event of a growing cell death, how robust are the pacemaker cells in the heart or the suprachiasmatic cells  in the brain to sustain a globally synchronized oscillation? In the dynamical sense, a death of a cell is considered as a passive or an excitable state. In the situation of a progressive  cell death, in other words, increasing number of passive oscillators, a population of globally coupled oscillators showed a type of aging transition \cite {daido}. Such  aging transition or death state is not the focus of this current study. We emphasize rather on the synchronized state (1:1 or higher phase-locking) of  global oscillation of the mixed population   as shown earlier \cite{daido, pazo} where the type of oscillatory  dynamics was not given appropriate attention. 
\par In this backdrop, we consider the superconducting RCSJ model to construct a globally coupled network of mixed population of excitable and oscillatory junctions and, particularly, focus on its collective coherent dynamics. Each individual junction is governed by the SNIC bifurcation to limit cycle oscillation. We distinguish the RCSJ units as excitable when they are in a stable steady state for a selected constant current bias less than a critical value and  oscillatory when biased by a higher constant current to cross the SNIC bifurcation point.  As a result, we find that the presence of a fraction of excitable units generates bursting in the whole network although the uncoupled oscillatory junctions  never show bursting dynamics. For a coupling above a threshold, the whole network starts synchronous firing with single spiking, and for further increase of coupling, periodic bursting appears with increasing number of spikes in a single burst and finally which clearly emerges as a parabolic bursting. During the spiking and bursting above a coupling threshold, the whole network splits into two synchronous clusters, one forming a synchronization manifold of the excitable units and another of the oscillatory units, however, they are phase-locked. We reduce the network model using the two synchronization manifolds of the oscillatory and excitable units and explain the  bursting mechanism  and furthermore, numerically verify the bursting dynamics of the whole network. 
\section{Single junction model}
 \par A single RCSJ model is described by,
\begin{eqnarray}
\label{Rcsj}
 \ddot{\theta}+ \alpha \dot{\theta} + sin\theta &=& I. 
 \end{eqnarray}
where 
  $\theta$ is phase difference of the junction, $\dot\theta$=$v$ is the voltage across the  junction,
   $\alpha$=$[h/2 \pi e I R^2 C]^{1/2}$ is the damping parameter, $h$ is the Planck's constant, $e$ is the electronic charge and $I$ is a constant bias current. 
  It has an equilibrium solution of sin $\theta$=$I_0$ in a cylindrical space. The stability of the equilibrium is obtained from the $f'(\theta^*)=cos \theta^*=±(1-I_0^2)^{1/2}$ where $f'$=$df/d \theta$ at equilibrium
$\theta=\theta^*$. For $I_0<1.0$, the model has clearly two equilibrium points, a node for $f'(\theta^*)<0$ and  a saddle
for $f'(\theta^*)>0$. They coalesce at $I_0=1.0$ via SNIC bifurcation \cite{levi, strogatz} for a choice of $\alpha>1.19$. 
For $\alpha<1.19$, a fold bifurcation is recorded at $I_0=1.0$. In addition there is a bistable region for $I_0<1.0$ and $\alpha<1.19$. 
 We focus here on the SNIC regime for $I_0>1.0$ and $\alpha>1.19$, where the stable equilibrium is separated from the oscillatory regime by a bifurcation line ($I_0=1.0$). 
 \begin{figure}
\includegraphics[width=9cm, height=9cm]{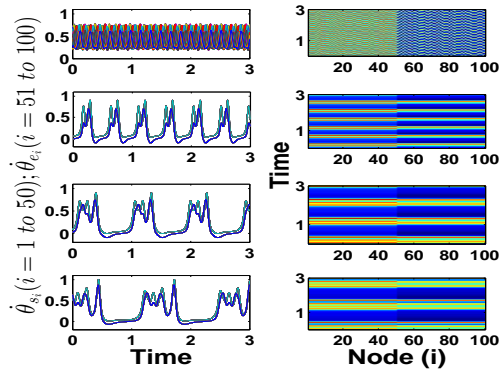}
 \caption{(Color online) Bursting dynamics in a network of Josephson junctions. Temporal dynamics shown in the left panels and spatio-temporal dynamics in the right panels for N=100. Fraction of excitable units  in the network, $p/N=0.5$. $I_e=0.5, I_s=1.5, p=0.5, \alpha=1.5$. Asynchronlous network for $\epsilon=3.7$ (panels in the uppermost row), two synchronous clusters with bursting  in the second row ($\epsilon=5.0$), third row ($\epsilon=8.0$),  bottom row ($\epsilon=9.7$). } 
\label{Spatio_temporal_network}
\end{figure} 
\section{Network of junctions}
 We consider a population of N globally coupled RCSJ units in which $p$ number of oscillators are in excitable mode $(I_{e_i}<1.0)$, in general, and $(N-p)$ units are self-oscillatory ($I_{s_i}>1.0$). The network consists of two subpopulations and its dynamics is described by two sets of equations,
\begin{eqnarray}
{\ddot\theta}_{e_i}+ \alpha_{e_i} {\dot\theta}_{e_i}+ sin\theta_{e_i} &=& I_{e_i}+\frac{\epsilon}{N}\sum_{j=1}^{N}({\dot \theta}_{s_j} - {\dot \theta}_{e_i}). \\  
{\ddot\theta}_{s_i}+ \alpha_{s_i} {\dot\theta}_{s_i} + sin\theta_{s_i} &=& I_{s_i}+\frac{\epsilon}{N}\sum_{j=1}^{N}({\dot \theta}_{e_j} - {\dot \theta}_{s_i}). 
\end{eqnarray}
where $e_i=1,2,...,p$ and ${s}_i=p+1,p+2,...,N$ denote the  excitable and self-oscillatory units respectively. 
The $\alpha=1.5$ is chosen identical for all the oscillators to restrict our current study in the SNIC regime \cite{levi, strogatz}. The bias currents to the excitable and oscillatory units are assumed as, $I_{e_i}=0.5$ and $I_{s_i}=1.25$ respectively.
\begin{figure}
\label{Network_bifur_osc}\includegraphics[width=8.5cm, height=8cm]{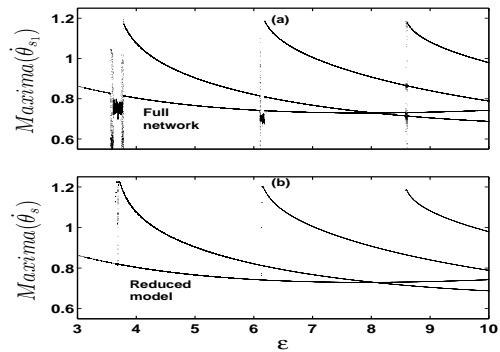}
\caption{Bifurcation diagram of a Josephson junction unit in the network. One oscillatory unit randomly chosen from the whole population and shown its bifurcation in the upper panel (a), and dynamics of the reduced model at lower panel (b). $I_e=0.5, I_s=1.5, \alpha=1.5$.} 
\label{maxima_x2_net_redu_model}
\end{figure} 
\par For numerical simulations, we first consider a network of size N=100 with equal number of oscillatory and excitable units. Initial conditions are chosen carefully using random numbers generated between 0.2 and 0.3. Figure \ref{Spatio_temporal_network} reveals a sequence of bursting oscillation in the whole network for increasing coupling strength in the upper to the lower panels except the uppermost panels. The panels in the uppermost row show no phase-locking for coupling strength $\epsilon=3.7$. For $\epsilon>3.7$ in rest of the panels, the  whole population forms two clusters as seen from the time series plot of all the oscillators ($\theta_{si}$, $s_i=50$ and $\theta_{ei}$, $e_i=50$) in each panel. The excitable and the oscillatory units form two separate clusters above a threshold coupling (panels in lower three rows), and the two subgroups are also seen phase locked. In fact, the first phase-locked firing in the whole network starts with single spiking dynamics (not shown here) above a coupling threshold  and then appears the bursting for larger coupling strength and adds on one after another spike in each burst (left panels in lower three rows). The number of spikes could be even larger for further increase of coupling strength as shown later, in the text, when we are able to recognize the parabolic nature of the bursting. Each of the right panels describes a temporal pattern of all the oscillator nodes ($s_i=50, e_i=50$); lower three panels clearly show formation of two clusters. These are in perfect match with the nature of the time series at their immediate left panels. This allows a reductionism approach \cite{daido, hens} to the large network dynamics and restrict them into two synchronization manifolds, $\theta_1=\theta_2=....=\theta_p$ representing the original excitable units and $\theta_{p+1}=\theta_{p+2}=....=\theta_N$ representing the original oscillatory units when we  represent the network by two oscillators,
\begin{eqnarray}
 \label{Rcsj}
 \ddot{\theta_e}+ \alpha_e \dot{\theta_e} + sin\theta_e &=& I_e+\epsilon (1-p) (\dot{\theta_s}-\dot{\theta_e}). \\ 
  \ddot{\theta_s}+ \alpha_s \dot{\theta_s} + sin\theta_s &=& I_s+\epsilon p(\dot{\theta_e}-\dot{\theta_s}). 
  \end{eqnarray}
where $p/N$ denotes the fraction of excitable junctions in the whole population. 
\begin{figure}
\includegraphics[width=9cm, height=4cm]{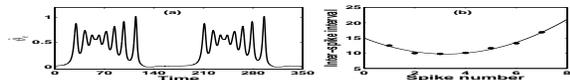}
\caption{{\it Circle/Circle} bursting in the network of Josephson junctions. Temporal dynamics in the left panel, inter-spike interval showing parabolic nature in the right panel.  Percentage of excitable units, $p=0.5$, $\epsilon$=$20$, $I_e=0.5, I_s=1.5, \alpha=1.5$.} 
\label{parabolic_bursting}
\end{figure} 
\begin{figure}
\includegraphics[width=7cm, height=5cm]{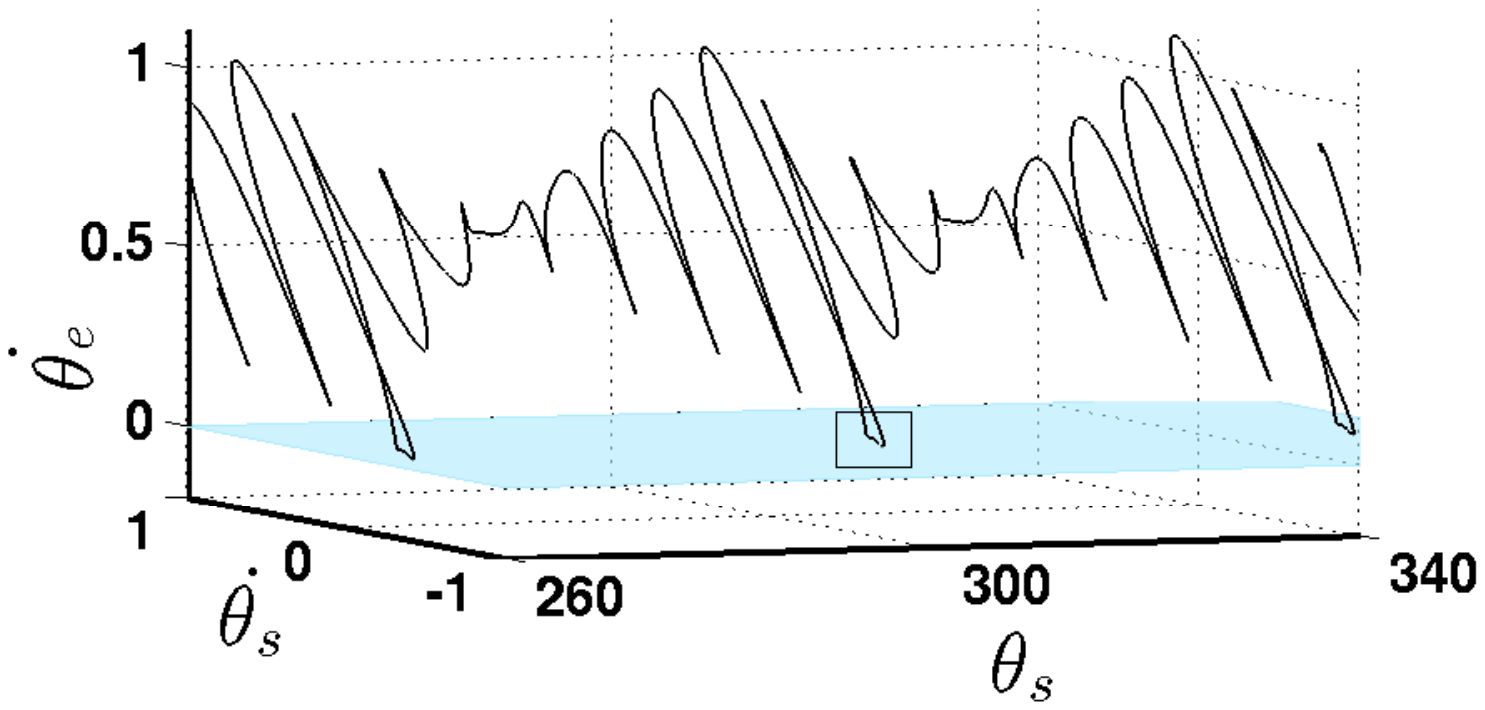}
\includegraphics[width=7.5cm, height=6cm]{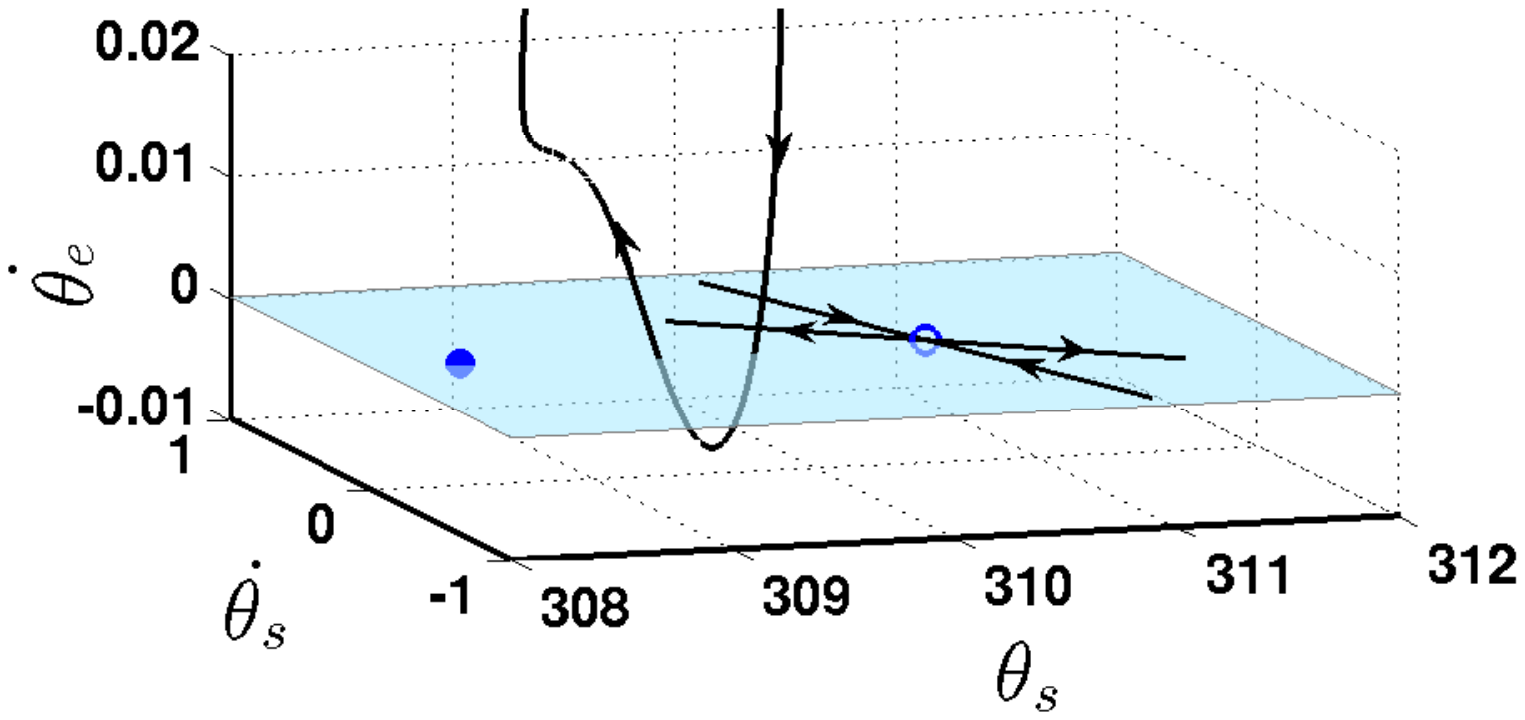}
\caption{(Color online) {\it Circle/Circle} bursting of the Josephson junctions. Upper panel shows the bursting in a 3D plane of $\dot \theta_e$, $\dot \theta_s$, $\theta_s$. Lower panel shows an enlarged picture of a part (in a box) of the upper panel. $p=0.5$, $\epsilon$=$20$, $I_e=0.5, I_s=1.5, \alpha=1.5$.} 
\label{3d}
\end{figure} 
\par Figure \ref{maxima_x2_net_redu_model} presents the bifurcation diagram of the dynamics of a single oscillatory unit arbitrarily chosen from the whole network and its reduced model (4)-(5) as well. Maxima of $\theta_{si}$ of the junction node (say, $i=1$) is plotted with coupling strength ($\epsilon$) in the upper panel which represents the original oscillatory units ($s_i$).  It shows periodic bursting with the number of spikes increasing in a burst one after another with coupling strength. Each period-adding regime is intercepted by a complex bursting window. The maxima of $\theta_s$ of the reduced model is shown in the lower panel and its bifurcation is in agreement with the upper panel. The windows of complex dynamics are also found matching, which also shows complex bursting pattern but here we do not focus on this feature here. The reduced model thereby perfectly represents the dynamics of the whole network.  
The excitable units ($e_i$) also show similar bifurcation diagram (not shown here) and match with the reduced model of the excitable units as expected since they are all phase-locked with the oscillatory units ($s_i$). 
\par The nature of  bursting is parabolic ({\it circle/circle} type) \cite{izhikevich, ermentrout} as shown in Fig. \ref{parabolic_bursting}, in the sense, that the oscillation stops via SNIC bifurcation and the trajectory moves towards the steady state that becomes unstable  via SNIC after an elapse of time to start the oscillation once again and the process repeats. Left panel shows the time series of few bursts only; a larger coupling strength is considered here, when the number of spikes is reasonably large. The inter-spike interval in a burst is plotted for sucessive spikes in the right panel, which confirmed the parabolic nature of the interspike intervals. 
The bursting dynamics periodically switches between the oscillatory state and the steady state both via SNIC bifurcation. The excitable units induces a slow dynamics in the individual units that controls the firing (oscillatory) and steady (resting) states which is further elaborated in Fig. \ref{3d}.
The upper panel demonstrates the {\it circle/circle} bursting in a 3D plane of $\dot \theta_e$, $\dot \theta_s$, $\theta_s$ where the zero plane is drawn in cyan/gray color. A part of the trajectory (in a box) goes below the zero plane which is enlarged in the lower panel. The location of the saddle with its eigen-directions and the node of the uncoupled excitable junction are denoted by open and soild circles respectively.
\par We further explain the bursting mechanism here as controlled by the slow dynamics. The trajectory of the bursting, after a few fast spiking,  moves towards the saddle (plane in cyan/gray line) guided by the stable manifold of the excitable unit. The trajectory obviously becomes slow when it approaches the saddle point, and after coming sufficiently close to it, moves away by the influence of the unstable egienvector. We make an approximation here that the excitable and the oscillatory units behave like isolated units since all the oscillators are now phase locked, when the error dynamics is negligbly small; the coupling term in the system is almost negligible. The ensemble of exctibale units and oscillatory  units are reduced to two junctions. The excitable unit maintains its isolated dynamics with a node and a saddle that influences the slow-fast bursting of the oscillatory junction. As a result, the network of junctions show a bursting dynamics typical of the Type I neuron. In fact, we observe the {\it circle/circle} bursting simply by coupling two oscillators, one excitable and another oscillatory.
\section{Conclusion}
\par In summary, we investigated a mixed population of oscillatory and excitable Josephson junctions under all-to-all global coupling when we observed {\it circle/circle} bursting in a broad parameter range of the junction and the coupling strength. We produced numerical evidence of the phenomenon using a network of N=100 oscillators and taking two equal populations of oscillatory and excitbale junctions.  The whole network splits into two clusters for our chosen range of coupling strength that helps reduce the system into a two-oscillator model. Results of the reduced model were found perfectly matching with the numerical results of the whole network. We found that the number of spikes increases with coupling strength which we supported with a bifurcation diagram of the whole network and its reduced model. The bursting dynamics had been a dominant feature of the mixed population such that it existed for different percentage of excitable units although we have only deatiled the case of fifty-fifty populations of oscillatory and excitable units. The bursting is typically {\it circle/circle} type in the selected  parameter regime of the superconducting device where the dynamics is governed by the SNIC bifurcation. We have numerically checked that the bursting also appeared  in a mixed population of another SNIC model, the Morris-Lecar system \cite{morris-lecar}. The bursting dynamics, therefore, seemed to be a generic feature of a mixed population of such dynamical units and of even a larger size. A mixed population of oscillatory and excitable units, in general, was investigated earlier by others \cite{daido, pazo}, however, they focussed on the global oscillation, partial oscillation and oscillation death  regimes. In the globally synchronized  (1:1 and higher phase locking) oscillation regime, the type of dynamics was  mentioned \cite{pazo} as could be of complex type, however, it was not focussed.  Our results, in that sense, provided additional information about the network dynamics of a mixed population in the global oscillation regime (1:1 phase locking) and, particularly, showed clear evidence of bursting in a SNIC model, the superconducting junction. Other bifurcation regimes of the junction such as the fold bifurcation and the bistable region are also interesting which we plan to explore, in the future, especially using other complex network topologies.

 \par The authors acknowledge Baruch Barzel for interesting discussions. C.R.H and S.K.D. acknowledge support by the CSIR Emeritus Scientist Scheme, India.

\end{document}